# What makes a complex liquid complex?

Alexander Z. Patashinski[a,b*], Rafal Orlik[c], Antoni C. Mitus[d], Mark A. Ratner[a,] and Bartosz A. Grzybowski[a,b*]

[a]Department of Chemistry, Northwestern University, 2145 Sheridan Road, Evanston, IL 60208-3113, USA;

[b]Department of Chemical and Biological Engineering, Northwestern University, 2145 Sheridan Road, Evanston, IL 60208-3113, USA

[c]Orlik Software, ul. Lniana 22/12, 50-520 Wrocław, Poland;

[d]Institute of Physics, University of Technology, Wybrzeze Wyspianskiego 27, 50-370 Wroclaw, Poland;

E-mails list:

Alexander Patashinski: a-patashinski@northwestern.edu

Rafal Orlik: rafal.orlik@orlik-software.eu

Antoni C. Mitus: antoni.mitus@pwr.wroc.pl

Mark A. Ratner: Ratner@chem.northwestern.edu

Bartosz A. Grzybowski: grzybor@northwestern.edu

*2D*-liquids near solidification are shown to be complex liquids characterized by a wide spectrum of relaxation times, stretched-exponential kinetics of sampling the equilibrium ensemble from a given configuration, and a power-law behavior of the distribution function $f(\tau_W)$ of the lifetimes $\tau_W$ of particles arrangements in small volumes of the liquid. Complexity arises from dynamical aggregation of particles: the equilibrium liquid is a dynamic mosaic of long-living crystallites separated by less-ordered regions. The long-time dynamics of this structure is dominated by particles redistribution between these dynamically and structurally different regions. We describe the microscopic details of this dynamics, and then discuss complexity in 2D and 3D liquids in the context of the general complexity paradigm.

Complex systems are distinguished by a rather general common feature: their behavior is determined by competing processes of self-organization (ordering) and self-disorganization (disordering) creating a hierarchical adaptive structure [1,2]. In mathematics, complexity is a quantifiable and measurable characteristic of this hierarchy. A notion of complexity is also used in amorphous materials exhibiting slow and non-exponential relaxation, in particular in glass-forming liquids and glasses. However in this case, complexity is not yet a quantifyible but rather a qualitative characteristic. Numerous experimental and theoretical studies and, more recently, computer simulations [ 3 - 18] revealed important macro-and mesoscopic details associated with materials complexity: dramatic slowing-down of structure changes on cooling, wide spectrum of relaxation times and stretched-exponential (KWW) relaxation kinetics [6], dynamic hererogeneity on microscopic length-scales [8-10]. These features and the sometime observed power-law correlations are often used as practical but rather qualitative criteria of complexity in materials. In the Literature, the assumed physical cause of materials complexity is the dynamic competition between aggregation of particles into preferred (essential [3], inherent [4]) structures, and factors preventing crystallization. Understanding the origins of complexity and the dynamics of structure in complex materials remains one of the most important but also hardest problems in condensed matter.

Not every liquid becomes complex on cooling. Three-dimensional (*3D*) liquids with simple two-particle interactions (molten metals and salts, liquefied noble gases, and also computer liquids of Lennard-Jones (*LJ*), soft core, Morse particles) aggressively crystallize on cooling before they show any significant signs of complexity. Classical 3D complex liquids have complicated and competing interactions, and special supercooling regimes are necessary to avoid crystallization on supercooling.

We note that, unlike the 3D case, two-dimensional (*2D*) liquids with simple interactions have [20,21] a continuous [22] or almost continuous [23] crossover from simple liquid state to crystal. At crossover temperatures (Fig.1), particles in these equilibrium liquids aggregate to form a dynamic mosaic [24,25] of crystalline-ordered regions (crystallites) and less-ordered clusters. At the high-temperature end of the mosaic states, crystallites are small and separated island of order in a disordered (amorphous) matrix. Crystallites fraction of the system increases at lover temperatures where crystallinity percolates. At even lower temperatures, crystallites merge into a multi-connected crystalline matrix with expected algebraic decay of orientation order (hexatic liquid) or long range order [20,22,23]. The mosaic is a feature observed at temperatures where the correlation length for orientations is finite and the 2D liquid is in normal (not hexatic) state.

The striking similarities of the above picture of *2D*-liquids at mosaic temperatures with that expected [5,10,16] in complex liquids motivated this study of complexity of 2D liquids in mosaic states (gray area in Fig.1). We found that the kinetics of structure changes in the liquid is stretched-exponential (Fig.3), lifetimes of local environments are power-law distributed (Fig.4), and space distribution of local changes is spatially-heterogeneous – typical features associated with complex liquids. However, while stretched exponents and power laws were observed in a wide range of temperatures, a developed multi-scale hierarchical system was found only in a narrow band of temperatures (dark-gray area *S* in Fig.1) where both order and disorder are in the regime of dynamical percolation. In this regime, the range of power-law behavior significantly increased while the values of the stretching exponent (Fig.3 and 4) decreased.

The intuitive idea of structure in a molecular system refers to nearest-neighborhood relations between particles. To define this structure as a coarsened characteristic of system configurations, we took advantage of the fact that in computer simulations, each particle is assigned a unique identifier $a=1, 2, …, N$, where $N$ is the particles number. A nearest neighbor (*nn*) of a particle $a$ is defined as one of $k$ particles neatest to $a$; in this study of *2D* liquids, $k=6$ is chosen to include all particles of the first coordination shell in a crystal. Connecting particles with their nearest neighbors by lines (bonds) creates a bond network, with particles as nodes and bonds as edges. A metric characteristic of the liquid at node $a$ is the bond orientation order parameter [26], a function of angles between bonds connecting the particle $a$ with its instantaneous nearest neighbors. This parameter was widely used to study 2D and 3D systems. Single bonds break-ups, defined as increase of the bond length beyond a set value, were studied [11-13] in *2D* and *3D* supercooled liquids and at melting.

To track changes in particles arrangements in the vicinity of a node $a$, we used here a different local characteristic - the list $L_a(t)$ (*nn*-list) of nearest neighbors of $a$. The *nn*-lists are topological characteristics of bond network connectivity, discrete functions of time conserved by small-amplitude vibrations. A change in $L_a(t)$ signals a rearrangement of particles in the cluster containing the particle $a$, a change in the bond network. Network changes are gradual: if the *nn*-lists of all particles are memorized at a time $t_0$, only a fraction $K(t)$ of these lists remains unchanged ($L_a(t_0)= L_a(t_0+t)$) at a time $t'= t_0+t$ .

In an equilibrium system, *K(t)* describes sampling of the equilibrium ensemble of structures starting from a chosen configuration.

In [11-13], a bond between two particles was declared brocken when the distance between these particles exceeded some set value. This definition is not directly based on topological features of the bonds network, but can be used in connection with these features. In recognizing structure changes, the two methods significantly differ for a large fluctuation of density in a small volume without changing the near neighbors, but coincide for nn-changes livig longer that particles vibration period. The physical picture emerging from [11-13] and this study is qualitatively similar; the main differences are in questions asked. The aim of our study is quantifying the dynamics of sampling the equilibrium ensemble of structures from a chosen configuration (in particular, the function *K(t)*), and searching for features related to complexity of this dynamics (in particular, the distribution of nn-lists lifetimes)-see below.

*2D* systems of *N*=587, 1024, 2500, and 16875 particles interacting via Lennard-Jones and soft core potential were simulated. Below, we mainly describe the data for a system of *N*=2500 *LJ* particles (interaction potential $U_{LJ}(r) = 4\varepsilon\left[(r_0/r)^{12} - (r_0/r)^6\right]$); results for soft-core particles were similar. For *LJ*-systems, thermodynamic states were scanned along the isochore $\rho^*$=0.84 and along the isotherm $T^*$= 0.70, where $T^* = k_B T/\varepsilon$ and $\rho^* = \rho r_0^2$, *T* and $\rho$ being the physical temperature and particles number density and $k_B$ the Boltzmann constant. We performed rather standard Molecular Dynamics (NVT) simulations (see details in [24,25]) under periodic boundary conditions with time step $h = \tau_{LJ}/500$ where $\tau_{LJ}$ is the *LJ* time unit [25], about the period of particles vibrations.

After equilibrating the system for at least $5*10^4$ $h$, configurations were saved at times $m\Delta t$, $m$=1, 2, …. , 2000, and $\Delta t=50h \sim \tau_{LJ}/10$ for analysis. Using the bond orientation order parameter as described in [25], we found that in the mosaic states (Fig.1), crystalline-ordered regions (henceforth, crystallites) occupied ~10% of the system at $T^*$=0.700, $\rho^*$=0.60, ~40% at $T^*$=0.700, $\rho^*$=0.80, and ~80% at $T^*$=0.700, $\rho^*$=0.86. Exact occupation values weakly depended on the recognition threshold [25] for crystallinity, but for any reasonable choice of this threshold the crystalline-ordered regions percolated (for $T^*$=0.70) in the range $\rho^*$=0.83-0.84. At $\rho^*$>0.86, crystallites merge in crystalline-ordered matrix with a small number of isolated defects (vacancies and dislocations).

For all states, functions $K(t)$ were calculated and fitted by the stretched-exponential function $K(t)=exp[-(t/\theta)^\beta]$ (inset in Fig. 3(a)); the best-fit values $\theta(T^*,\rho^*)$ and $\beta(T^*,\rho^*)$ are plotted in Fig.3. Over the scanned range of states, the *nn*-relaxation time $\theta(T^*,\rho^*)$ changed by five orders of magnitude (Fig.3(a)). At the high-temperature/low density end, $\theta(T^*,\rho^*)$ was commensurate with $\tau_{LJ}$, and simulation time was $200\tau_{LJ}$; at temperatures where the system was almost crystalline, $\theta(T^*,\rho^*)$ exceeded $10^4\tau_{LJ}$, and runs with an extended simulation time of $10^4\tau_{LJ}$ were performed. Fig. 3(a) shows the Arrhenius plot for the isochore $\rho^* = 0.84$. In the mosaic range (Fig.1), this plot deviates from linear and corresponds to a modestly fragile [5] glassformer. At higher densities crystallites percolate into a crystalline matrix, *nn*-changes are produced by moving defects, and the Arrhenius plot becomes linear.

Along the isotherm $T^*$=0.70, the stretching exponent has a minimum $\beta_{min} \approx 0.5$ at $\rho^* \approx 0.84$, and $\beta$<0.60 for $\rho^* = 0.83$-0.85 (dark-gray area $S$ in Fig. 1). For $T^*$=0.70 and $\rho^*$<0.84, value of $\beta$ fluctuated between runs: $\Delta\beta= \pm 6\%$; these fluctuations increased with increasing density. At $\rho^*\geq 0.86$, the functions $K(t)$ for each run substantially deviated from each other and from the fit curve.

Color-coding particles when they change their *nn*-lists (Fig.2) revealed the spatial heterogeneity of the rates of *nn*-changes. Dynamic heterogeneity is a known feature of *3D* glassformers, conformed by experiments using molecules-sized probes imbedded in the material [2,10,5], and computer simulations [11-15]. In *2D* liquids studied here, the dynamic heterogeneity was strongly correlated with structural heterogeneity of the mosaic: *nn*-changes were only detected outside or at the borders of crystallites (see Fig.2). While particles inside a crystallite only vibrated without rearrangements, an *nn*-change at their border frequently resulted in a small (1-3 particles) local increase or decrease of the crystallite. On times $\sim\theta(T^*,\rho^*)$, this local process of micro-crystallization/micro-melting substantially changed positions, shapes, and orientations of crystallites, so the space distribution of rigid and soft regions had its own dynamics. In equilibrium systems studied here, this dynamics defines how the system samples the equilibrium ensemble starting from a chosen configuration; the stretched-exponential kinetics of this sampling is described by the function $K(t)$.

The use of *nn*-lists reduces the description of particles continuous motions to two cases: either waiting (vibrations conserving the *nn*-list) for a time $\tau_W$ without changing the *nn*-list, or changing to a new list. We found that the waiting times $\tau_W$ are distributed over a wide range of times, from very small to times exceeding the nn-relaxation time $\theta(T^*,\rho^*)$. The *log-log* plot of the waiting times distribution function $f(\tau_W)$ is shown in Fig.4. For all temperatures and densities scanned, the plot is approximately linear in the range $0.5\tau_{LJ} < \tau_W < \tau_{Wc}$, indicating a power law $f(\tau_W) \sim \tau_W^{-\beta_W}$. At $T^* = 0.700$, $\rho^* = 0.78$, $\tau_{Wc} \sim 5\tau_{LJ}$, and $\beta_W \sim 1.5\pm 0.2$; at $T^* = 0.700$, $\rho^* = 0.86$, $\tau_{Wc} \sim 10\tau_{LJ}$, and $\beta_W = 1.8 \pm 0.2$.

Short ($\tau_W \ll \theta(T^*,\rho^*)$) and longer waiting times describe different types of events. The power law distribution at small $\tau_W$ describes frequent rearrangements in the areas outside of crystallites. Many of these short-living *nn*-changes appear reversible: after a short waiting time, the local bond network returns to its pre-rearrangement state. The long waiting times characterize particles trapped inside crystallites. These particles keep nearest neighbors until the moving crystallite's border penetrates the crystallite to free them. Thus, a large $\tau_W$ includes the time of continuously being inside a crystallite. Observation of particles diffusion supports this interpretation: a particle performes many vibrations without much displacement until it becomes highly mobile and makes a series of large diffusion steps before becoming inactive again; this heterogeneous diffusion was observed in experiments [5,8,9,10] and simulations [11,12,15] in supercooled liquids and glasses.

Between high-temperature (mostly amorphous) and low-temperature (mostly crystalline) states, there is a narrow range of states (the dark-gray area *S* in Fig.1) in which the crystalline part of the material and the part not recognized as crystalline occupy approximately approximately equal parts of the system, and both are close to percolation (see Fig.2). In equilibrium systems studied here, percolation is dynamic [29]: at any time, the bonds connecting particles recognized as crystalline form a multy-connected fractal cluster, but these clusters are different at different times. In percolation states, the power-law for $f(\tau_W)$ extends to waiting times much larger than outside *S*; on the isotherm $T^*=0.700$, the time $\tau_{Wc}(T^*,\rho^*)$ has a pick at $\rho^*\approx0.84$ where $\tau_{Wc}\sim100\tau_{LJ}$. Inside **S**, also the stretching exponent $\beta(T^*,\rho^*)$ has a dip with a minimum.

In mathematical models of percolation [28,29], the percolating cluster in an infinite system is a complex multi-scale hierarchical system. The fractal, scaling nature of this cluster leads to power laws for space correlations at distances limited from below by the smallest scale in the hierarchy $R_{min}$ set by the construction of the system, and from above by the largest scale (the percolation length) $R_{max}$ that is a function of controlling parameters (for example, the fraction of broken bonds in bond percolation on a lattice). The number of scales in the hierarchical system can be defined as

$$M = \ln\frac{R\max}{R\min} .$$    **Equation 1**

As a function of controlling parameters, M has a singularity at percolation point, smoothed to a maximum ~ln(L/ Rmin) in a system of a size L.

Rigidity percolation as the cause of "dynamic heterogeneity and its sensitivity of complex behavior to changes in bond configuration" in glassformers was recently demonstrated in [30]. Unlike the *2D LJ* liquid, generally the order in rigid agglomerates is variable and not easily recognizable, although, as one suggests, there always exist correlation between local structure and mobility [16,30,31]. In equilibrium liquids of which the *2D LJ* liquid is an example, the volume fraction of rigid agglomerates and the morphology of the rigid clusters is a function of state, and rigidity percolates at low temperatures. In supercooled or otherwise inherently unstable systems, the volume fraction of rigidity, and the morphology of the rigid clustera is a function of time and cooling history. In all cases, percolation can be assumed as the cause of a developed hierarchical structure in complex liquids. This relation of complexity to rigidity percolation allows one to quantify complexity of liquids using the number $M$ of scales in the hierarchy as the simplest measure. Depending on how close the system is to percolation, it may be more or less complex.

In *2D* liquids studied here, complexity is low ($M(T^*,\rho^*) \sim 1$) outside the percolation band of states *S*, although some features associated with complexity (stretched-exponential relaxation, and power-law distribution of waiting times) are already present. Inside the band *S* where $M(T^*,\rho^*)$ increases; here, for each isotherm there is a density $\rho^*_{min}(T^*)$ where $M(T^*,\rho^*)$ has a maximum. Within the accuracy of our data, at $\rho^*=\rho^*_{min}(T^*)$ also the stretching exponent $\beta(T^*,\rho^*)$ has a minimum and the characteristic time $\tau_{Wc}$ a maximum.

The finite size of simulated systems limits the lengh $R_{max}$, and thus the maximum complexity of the liquid. $R_{min}$ is about the size of the smallest crystallite, ~3 reduced units. Complexity is low ($M(T^*,\rho^*)= M_0 \sim 1$) beyond *S*. The maximum value of $M(T^*,\rho^*)$ is about $M_0+\Delta M$, with $\Delta M$=2,4,5,5.5 for systems of $N$=587, 2500, 16875, 60000 particles; simulation of even a 2DLJ system with complexity, for example, 100 is a challenging task. To compare properties change with increasing size, we simulated systems of $N$=587, 1024, 2500, and 16875. At temperatures and densities outside the percolation band of states, all these systems had qualitatively-similar features of the mosaics, and values of stretching exponent $\beta(T^*,\rho^*)$ coincided within the accuracy of our data. Inside the percolation band of states, larger systems had smaller values of $\beta(T^*,\rho^*)$ as shown in Fig. 3(b).

The dynamic percolation observed here is special: the power law in Fig. 4 indicates that in the time interval ($\tau_{W,min}$ - $\tau_{W,max}$) the liquid has a hierarchy of relaxation times, characterized by its own complexity $\tilde{M} = \ln\frac{\tau_{W,max}}{\tau_{W,min}}$. This feature is absent in models of dynamics where the probability of a local change is independent from the configuration of crystallites, and can only be explained by a feed-back interaction synchronizing local changes changes and changes in crystallites shapes and orientations. We are not aware of any theoretical or experimental studies of this new and very interesting situation of critical dynamic percolation, and have yet no quantitative explanation of the new scaling lows found in the *2D LJ* liquid. Although finite-size effects in complex systems are not expected to be large outside the percolation states [32], to answer the most interesting question if percolation in the mosaic extends to arbitrary large scales and leads to dynamic singularities, one needs to study systems of much larger sizes than those simulated in this study.

Conclusion:

- We found in *2D* liquids typical features commonly observed or expected in complex liquids: stretched-exponential relaxation, power-laws, and a dynamic mosaic of long-living aggregates separated by regions with short-living local structure. Changes in long-living aggregates are due to motions of their borders slowly changing the mosaic changes in a regime of dynamic percolation.
- In a narrow band of states in the phase plane, the balance between order and disorder creates a regime of critical dynamic percolation of long-living aggregates. Here, the stretching exponent has a minimum and power-law extends to large times and has a maximum.

**Acknowledgments:** This work was supported by the Nonequilibrium Energy Research Center (NERC) which is an Energy Frontier Research Center funded by the U.S. Department of Energy, Office of Science, Office of Basic Energy Sciences, under Award Number DE-SC0000989.

*References*

1. F. Heylighen, J. Bollen, and A. Riegler, Eds., *The Evolution of Complexity* (Kluwer, Dordrecht, 1999).

2. M. E. J. Newman, The physics of networks. Phys. Today **61**, 33–38 (2008).

3. H. Eyring, J. Chem. Phys. **4**, 283 (1936).

4. F. H. Stillinger and T. A. Weber, Science **225** , 983 (1984).

5. C. A. Angell, MRS Bulletin **33**, 544 (2008).

6. J. C. Phillips, Rep. Prog. Phys. **59,** 1133 (1996).

7. See reviews in: Science **267** (1995); *Supercooled Liquids: Advances and Novel Applications*, ACS Symposium Series 676, edited by J. T. Fourkas, D. Kivelson, U. Mohanty, and K. A. Nelson ~American Chemical Society, Washington, D.C., 1996; *Jamming and Rheology: Constrained Dynamics on Microscopic and Macroscopic Scales*, edited by A. Liu and S. R. Nagel, Taylor & Francis, London and New York, 2001.

8. L. A. Deschenes, D. A. Vanden Bout, Science **292***,* 255 (2001).

9. M. D. Ediger, Annu. Rev. Phys. Chem. **51***,* 99 *(*2000*)*.

10. M. D. Ediger and J. L. Skinner**,** Science,**292**,. 233-234 (2001)

11. H. Shiba, A. Onuki and T. Araki, EPL, **86** 66004 (2009)

12. ; R. Yamamoto and A. Onuki, Phys. Rev. Lett. **81**, 1415-1418, (1998).

13. A. Widmer-Cooper, H. Perry, P. Harrowell, and D. R. Reichman, Nature Physics **4**, 711-715 (2008).

14. P. G. Debenedetti and F. H. Stillinger, Nature (London) **410**, 259 (2001).

15. K. E. Avila, H. E. Castillo, A. Parsaeian, Phys. Rev. Lett. **107**, 265702 (2011).

16. V. Lubchenko, and P. G. Wolynes, Annu. Rev. Phys. Chem. 58, 235–266 (2007).

17. L. Berthier and G. Biroli, Rev. Mod. Phys. 83, 587–645 (2011).

18. H. Tanaka, T. Kawasaki, H. Shintani, & K. Watanabe, Nature Mat. 9, 324–331 (2010).

19. P. Ronhovde, S. Chakrabarty, D. Hu, M. Sahu, K. K. Sahu, K. F. Kelton, N. A. Mauro, & Z. Nussinov, Scientific Reports **2**, Article 329, 2012.

20. B. I. Halperin, D. R. Nelson, Phys. Rev. Lett. **41**, *121 (1978)*.

21. K. J. Strandburg, Rev. Mod. Phys. **60**, 161 (1988).

22. : K Bagchi; H. C. Andersen; W. Swope , Physical Review E **53,** 3794-3803. 1996.

23. K. Chen, T. Kaplan, and M. Mostroller, Phys. Rev. Lett. **74**, 4019 (**1995)**.

24. A. C. Mitus, A. Z. Patashinski, A. Patrykiejew, and S. Sokolowski, Phys. Rev. B **66**, 184202 (2002).

25. A. Z. Patashinski,R. Orlik, A. C. Mitus, B. A. Grzybowski, and M. A. Ratner, J. Phys. Chem. *C***, 114***,* 20749 (2010**)**.

26. P. J. Steinhardt, D. R. Nelson, & M. Ronchetti, Phys. Rev. B 28, 784–805 (1983).

27. B. I. Halperin, D. R. Nelson, Phys. Rev. Lett. **41**, *121 (1978).*

28. D. Staufer and A. Aharony, Introduction to Percolation Theory, (Taylor & Francis, 1994).

29. S. D. Druger, A. Nitzan, and M. A. Ratner, J. Chem. Phys. **79**, 3133 (1983).

30. V. K. de Souza and P. Harrowell, PNAS **106**, 15136–15141, (2009).

31. G. Tarjus, S. A. Kivelson, Z. Nussinov, and P. Viot, J. Phys.: Cond. Matter **17**, R1143 (2005).

32. L. Berthier, G. Biroli et all, *arxiv.org/pdf/1203.3392*

## *Figures Captions*

**Figure 1.**

The ($T^*$-$\rho^*$) thermodynamic plane: gray area − Mosaic states with 15% -80% of particles in crystallites, dark grey (*S*) – states where the crystallites percolate, the power-law for waiting times extends over the whole simulation interval (see text), and the stretching exponent is below the 0.65 value. The isotherm $T^*$=0.70 and isochore $\rho^*$=0.84 shown.

**Figure 2.**

The Mosaic and *K(t)* picture for $T^*$=0.70, $\rho^*$=0.84 at a time *t*=60 after the *nn*-lists were memorized at *t* = 0. Color coding: black – particles having $L_a(0)=L_a(60)$), blue-particles in new crystallites formed by micro-melting/crystallization (see text), red – particles that changed their *nn*s shortly ($1\tau_{LJ}$) before $t=60$, green – particles in amorphous regions.

**Figure 3.**

(a) Arrhenius plot for the relaxation times; inset: an example of a stretched-exponential fit for the function $K(t)$ ($T^*=0.70$, $\rho^*=0.84$).

(b) The values of the stretching exponent $\beta(T^*,\rho^*)$ along the isotherm $T^*=0.70$. Selected error bars shown

**Figure 4.**

*Log-log* plots of the waiting times distribution function $f(\tau_W)$ normalized by the total number $N_W$ of *nn*-changes. Main plot: $T^*=0.70$, $\rho^*=0.85$, inset: $T^*=0.70$, $\rho^*=0.81, 0.86$.

*Figures*

**Figure 1**

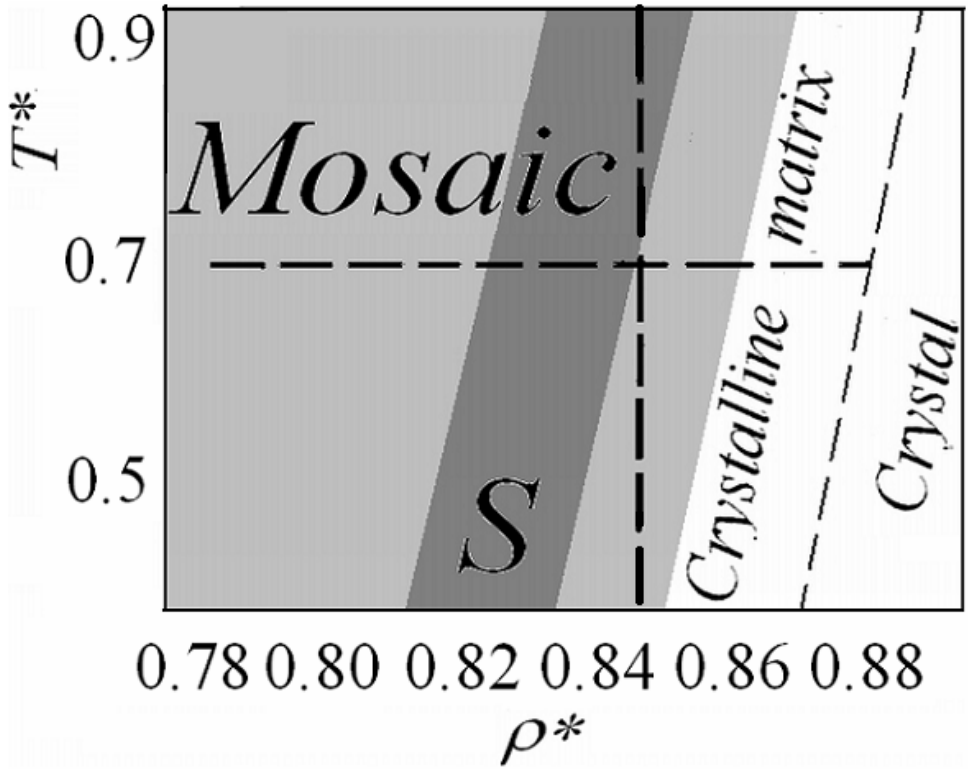

**Figure 2**

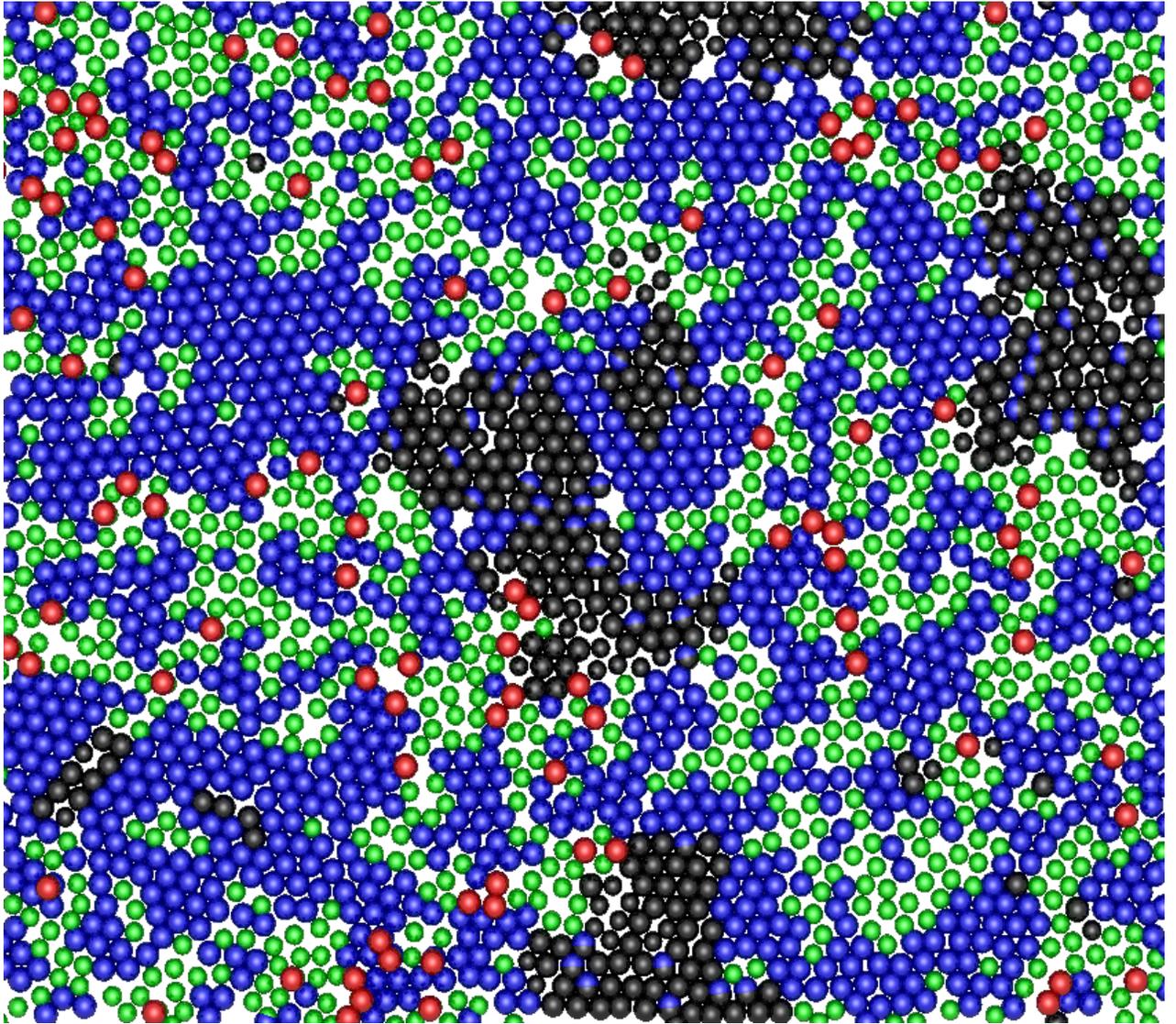

**Figure 3**

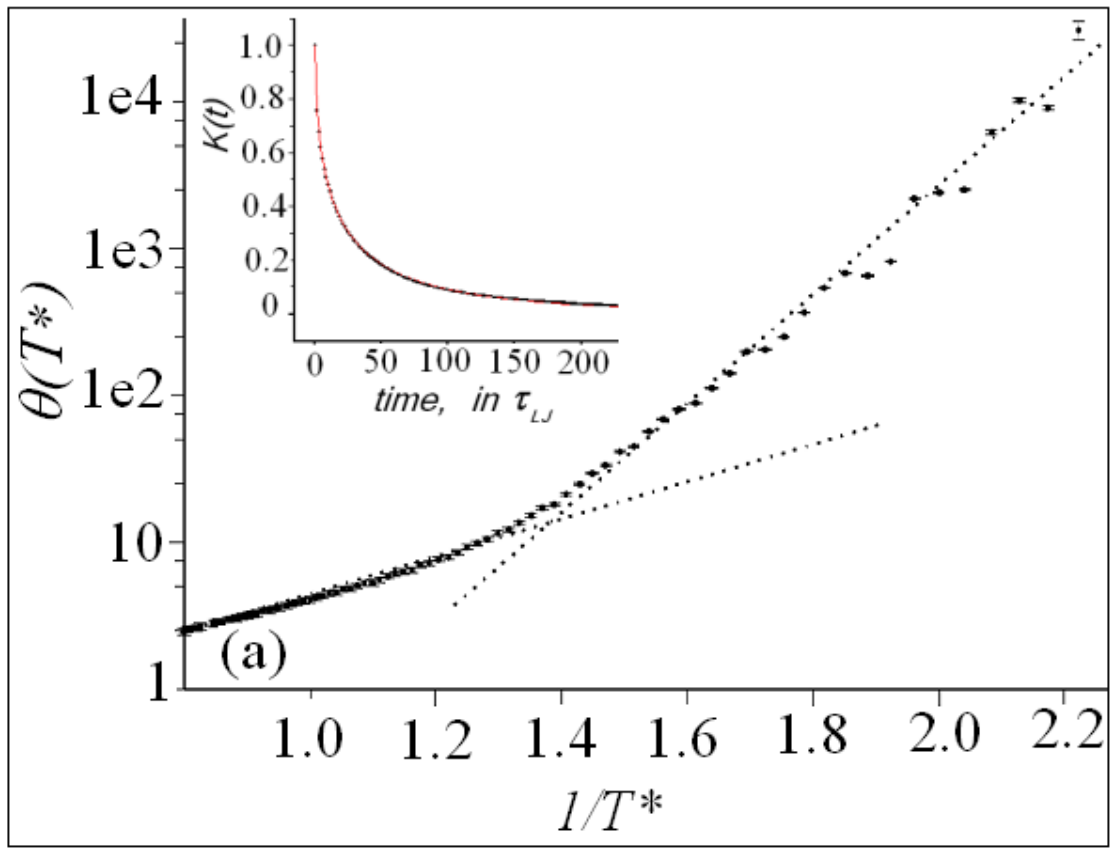

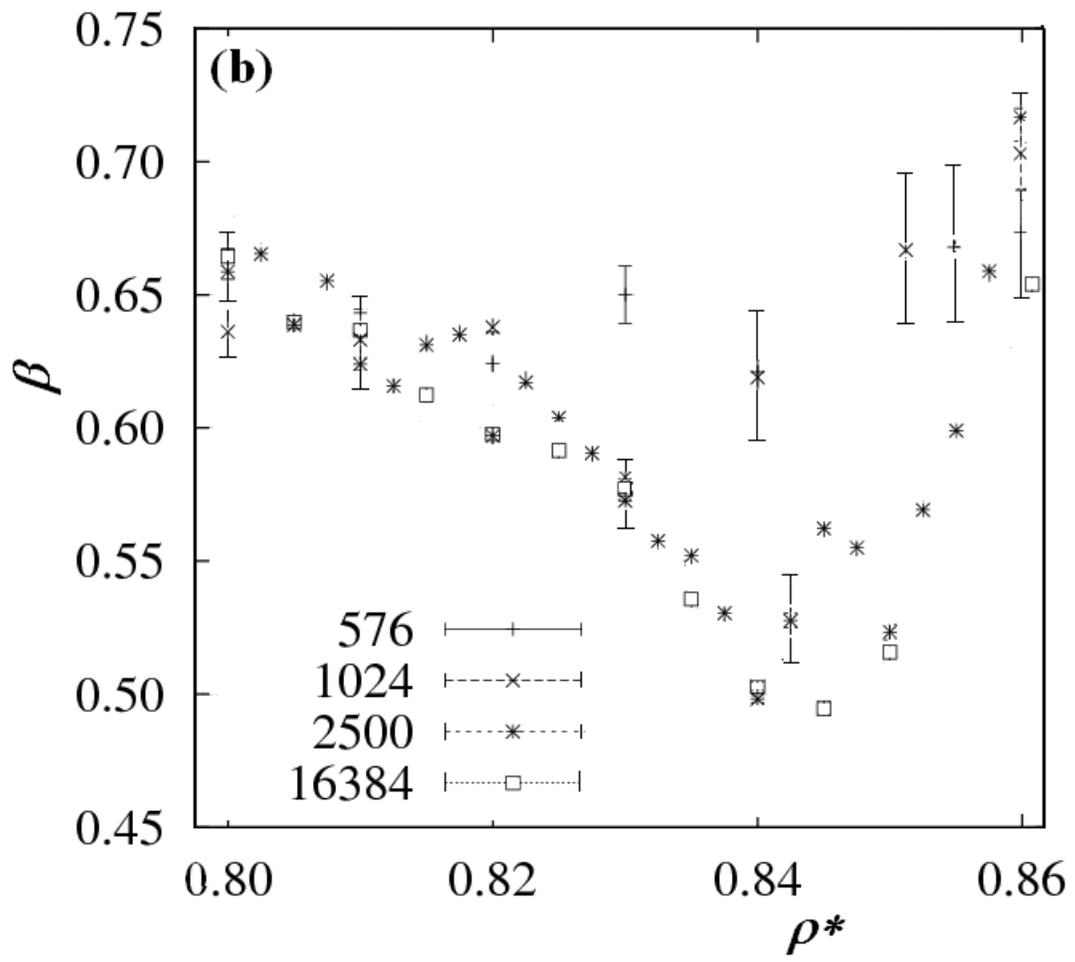

**Figure 4**

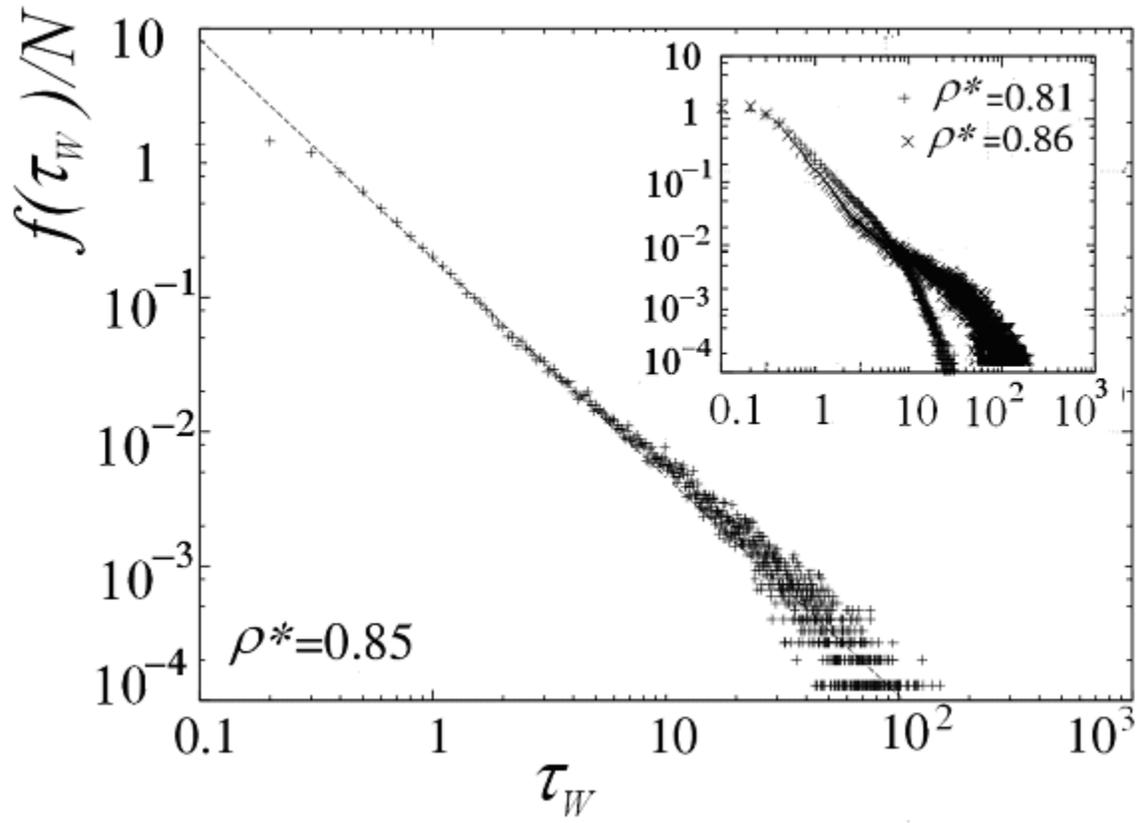